\documentclass[twoside,leqno,twocolumn]{article}

\usepackage[letterpaper]{geometry}
\usepackage{ltexpprt}
\usepackage{hyperref}
\usepackage{graphicx}
\usepackage{subcaption}
\usepackage{enumitem}
\usepackage{amsmath}
\usepackage{breqn}
\usepackage{lipsum}
\usepackage{multirow}
\usepackage[normalem]{ulem}
\usepackage{caption}
\usepackage{tabu}
\usepackage{multibib}
\usepackage{algorithm}
\usepackage{algpseudocode}
\usepackage[T1]{fontenc}
\usepackage{cancel}
\usepackage{xcolor}
\usepackage{titlesec}

\usepackage{array}
\usepackage{adjustbox}
\usepackage{mathtools}
\usepackage{cellspace}
\usepackage{float}
\usepackage{mathrsfs}

\usepackage[noadjust]{cite}

\newtheorem{definition}{Definition}

\usepackage{etoolbox}
\usepackage{titlesec}

\begin{document}

\title{Towards Spatially-Lucid AI Classification in Non-Euclidean Space: An Application for MxIF Oncology Data \footnote{This is the full version of the paper in SIAM DM 24.}
}

\author {Majid Farhadloo \thanks{Dept of Computer Science and Eng, University of Minnesota \{farha043, sharm485, gupta423, shekhar\}@umn.edu.}
\and Arun Sharma \footnotemark[2]
\and Jayant Gupta \footnotemark[2]
\and Alexey Leontovich \thanks{Mayo Clinic \{leontovich.alexey, markovic.svetomir\}@mayo.edu} 
\and Svetomir N. Markovic \footnotemark[3]
\and Shashi Shekhar \footnotemark[2]
}
\date{}

\maketitle

\fancyfoot[R]{\scriptsize{Copyright \textcopyright\ 2024 by SIAM\\
Unauthorized reproduction of this article is prohibited}}

\begin{abstract} \small\baselineskip=11pt
Given multi-category point sets from different place-types, our goal is to develop a spatially-lucid classifier that can distinguish between two classes based on the arrangements of their points. This problem is important for many applications, such as oncology, for analyzing immune-tumor relationships and designing new immunotherapies. It is challenging due to spatial variability and interpretability needs. Previously proposed techniques require dense training data or have limited ability to handle significant spatial variability within a single place-type. Most importantly, these deep neural network (DNN) approaches are not designed to work in non-Euclidean space, particularly point sets. Existing non-Euclidean DNN methods are limited to one-size-fits-all approaches. We explore a spatial ensemble framework that explicitly uses different training strategies, including weighted-distance learning rate and spatial domain adaptation, on various place-types for spatially-lucid classification. Experimental results on real-world datasets (e.g., MxIF oncology data) show that the proposed framework provides higher prediction accuracy than baseline methods.\\
\textit{\textbf{Keywords:} spatially-lucid, non-euclidean space, spatial variability, explainability, tumor oncology, MxIF}
\end{abstract}

\maketitle

\section{Introduction}
Given multi-category point sets from a non-Euclidean space (e.g., cellular spatial maps) from different place-types (e.g., tumor regions), our goal is to develop a spatially-lucid classifier. This approach is not only concerned with predicting accurate class labels but also emphasizes the need for interpretation \cite{whAIbill2023} in decision-making based on the spatial arrangements of data points. Spatial variability presents a significant challenge, wherein patterns and arrangements indicative of a class in one domain might not apply in another. For instance, spatial domain $I$ distinguishes between class $1$ and class $2$ using the arrangements of <\textcolor{red}{circle} and \textcolor{blue}{triangle}>. However, due to spatial variability, this same pattern does not reliably distinguish class labels in the Spatial Domain $II$. Instead, the distinct arrangement is the three-way configurations of <\textcolor{red}{circle}, \textcolor{blue}{triangle}, and \textcolor{green}{square}>. An equally paramount challenge lies in ensuring spatial explainability. Our proposed method is spatially-explainable since it is built based on the spatial arrangements of data points and help identify most discriminative features. This is achieved through post-hoc explainable methods (e.g., feature permutation) \cite{gunning2019xai} which delve into the model's decision processes to offer insights into which spatial arrangements are most influential in driving a particular classification outcome. This work focuses on developing a spatially-lucid classifier to address the challenges posed by spatial variability and explainability observed in non-Euclidean spaces. 



\begin{figure}[h]
    \centering
    \includegraphics[width=0.9\linewidth]{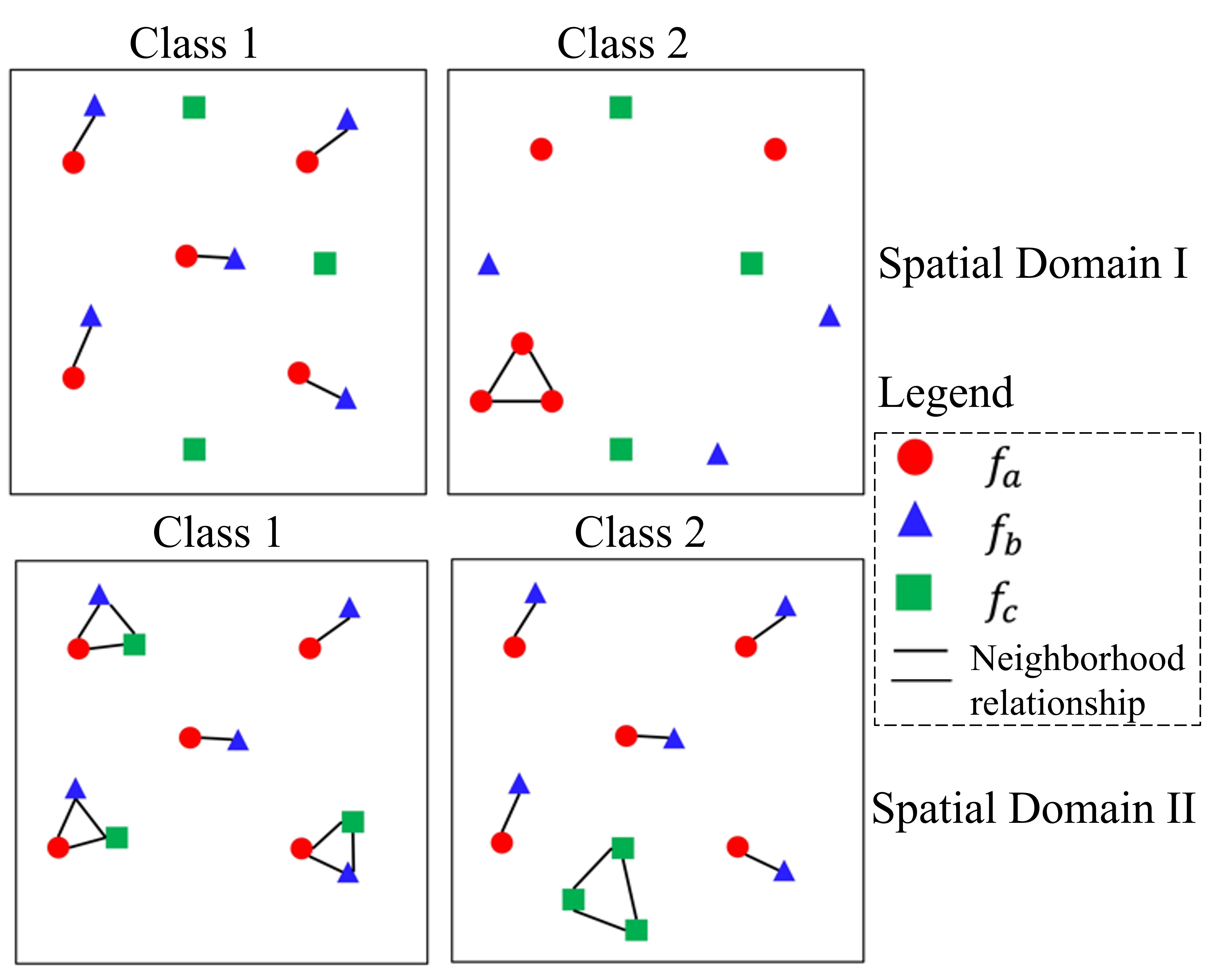}
    \caption{An illustration of the spatial differences in arrangements between two class labels of multi-category point sets from different spatial domains.}
    \label{input_output}
\end{figure}


Spatial variability is a prominent feature exhibited by many phenomena. Examples include language, cultural events, and even electronic circuits. Factors such as voltage (e.g., 120 voltage, 220 voltage), frequency, and plug types exemplify the variations observed across different spatial regions, spanning cities, countries, and continents such as North America and Europe. Harnessing spatial variability has proven instrumental in optimizing resource management and interventions in agriculture for maximizing crop yield \cite{paccioretti2020fastmapping}. In bio-medical applications, spatial variability also plays an important role. Neuroscientists study the spatial patterns of brain activity, enabling them to map cognitive processes and identify distinct brain networks involved in functions such as language and memory \cite{townsend2018detection}. Section \ref{application_domain} describes an illustrative bio-medical application domain showcasing the significance of spatial variability. 
Therefore, understanding spatial variability is crucial for comprehending spatial patterns and phenomena in diverse contexts.

Previously, we developed a spatial-variability-aware neural network (SVANN) \cite{gupta2020towards,gupta2021spatial} that uses location-dependent weights and found that it could better model variability than one-size-fits-all (OSFA) approaches. Other methods \cite{xie2021statistically, xie2021spatial} learn space partitions of the heterogeneous data and develop models tailored to the resulting homogeneous regions. However, these approaches either require dense training data for each spatial domain or struggle when significant variability exists within a domain. Most importantly, these techniques are based on traditional convolutional neural networks (CNNs), which are arranged on regular pixel grids or sequence, thus do not work in non-Euclidean space. As point cloud data has gained popularity, techniques for learning representations of point sets have attracted more attention \cite{qi2017pointnet, wang2019dynamic, zhao2021point}. These techniques work in non-Euclidean space, but do not fully leverage spatial relationships between multi-category points. Current methods mainly focus on point sets with few numerical attributes, such as signal strength, and they do not handle categorical attributes. To be more specific, these methods predominantly focus on per-point feature extraction without operations to effectively encode local geometric relationships and interactions between points of different categories in the neighborhood context. We recently developed a spatial-interaction aware multi-category deep neural network (SAMCNet) \cite{farhadloo2022samcnet} to represent spatial relationships and explore larger subsets of point types. However, SAMCNet and similar approaches (e.g., \cite{guo2020deep}) does not model spatial variability, since they use scalars rather than map weights.

To overcome these limitations, we explore a spatial ensemble framework that explicitly uses different training strategies with different place-types, including weighted-distance learning rate and spatial domain adaptation for spatially-lucid classification.\\
\textbf{Our contributions are as follows:}
\begin{itemize}[noitemsep,topsep=0pt]
    \item We deploy a spatial ensemble framework for spatially-lucid classification, where the network parameter is a map varying across place-types rather than a scalar as used in traditional OSFA approach.
    \item We demonstrate training strategies such as adjustable learning rates and spatial domain adaptation layers to address insufficient learning samples.
    \item Experiments show the proposed model outperforms existing baseline methods.
    \item A case study highlights the impact of spatial variability on tumor classification, illustrating cellular interactions that span from general (place-type independent) to specific (place-type dependent) interactions. This aims to enhance the manual assessments provided by pathologists.
\end{itemize}

\textbf{Scope:} 
This paper focus on spatial-interaction and variability-aware AI for multi-category point sets, targeting spatially-lucid classification. As these data inherently function as permutation-invariant set operators lacking predefined connectivity like graphs, comparisons with graph CNNs \cite{zhang2019graph} for structured data are out of scope. Similarly, comparisons with traditional CNNs using regular grid images as input are beyond the scope. Expanding the proposed methods to tackle high-dimensional variability is beyond our scope. We refrain from publishing the dataset due to patient privacy. We made the code used in the experiments available through GitHub \footnote{https://github.com/majid-farhadloo/Non-Euclidean-Space-SpatialEnsemble.git}.

\textbf{Organization:} The rest of the paper is organized as follows.  Section \ref{application_domain} briefly describes application domain of this problem. Related work is reviewed in Section \ref{related_work}. Section \ref{sec:Problem} introduces key concepts and formally defines the problem. Our proposed methods are described in Section \ref{sec:ProposedApproach}. Section \ref{sec:Experiment} presents the evaluation of the proposed method, followed by a case study in Section \ref{case-study}. Section \ref{sec:Conclusion} concludes the paper and outlines future work.

\section{An Illustrative Application
Domain}\label{application_domain}
In cancer research, understanding tumor interactions with normal tissues is essential for progression insights and immune therapy development. Multiplexed immunofluorescence (MxIF) imaging, especially relevant for immunocheckpoint inhibitor therapy (ICI), offers a detailed cellular spatial map of tumor and immune cells. Fig. \ref{fovs}, shows a map of different cell types (e.g., tumor and immune cells) with their corresponding locations. Although ICI therapy targets cancer cells by activating specific T lymphocytes, its effectiveness depends on complex \textit{spatial arrangements} within the tumor microenvironment (TME) \cite{andreou2022multiplexed, li2022cscd}.

\begin{figure}[h]
    \centering
    \includegraphics[width=0.85\linewidth]{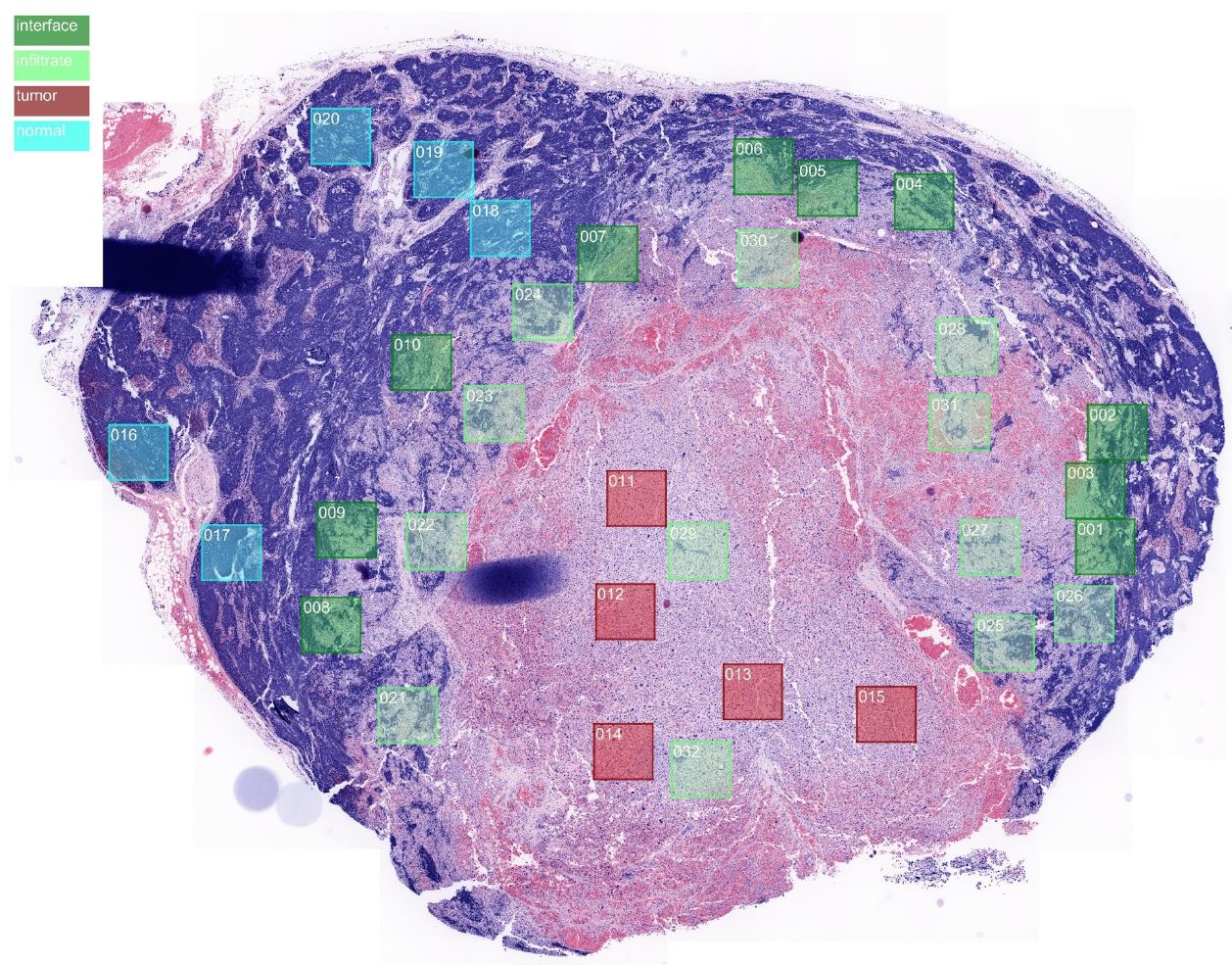}
    \caption{Spatial variability in a tissue slide, emphasized by pathology-driven fields of view (FOVs, colored squares), reveals distinct patterns within each FOV.}
  \label{h&e}
\end{figure}

While recent studies \cite{nawaz2016computational} offer insight into TME heterogeneity, modeling its spatial variability is challenging due to factors such as rapid cancer cell proliferation, genetic instability, and the presence of unknown mediator cells with immune and target cells. Figure \ref{h&e} displays spatial variability across a tissue slide using three colored anatomic structures. Our method aims to algorithmically describe these spatial patterns, potentially improving pathologists' visual assessments.

\section{Related Work}\label{related_work}
\begin{figure}[h!]
    \centering
    \includegraphics[width=0.65\linewidth]{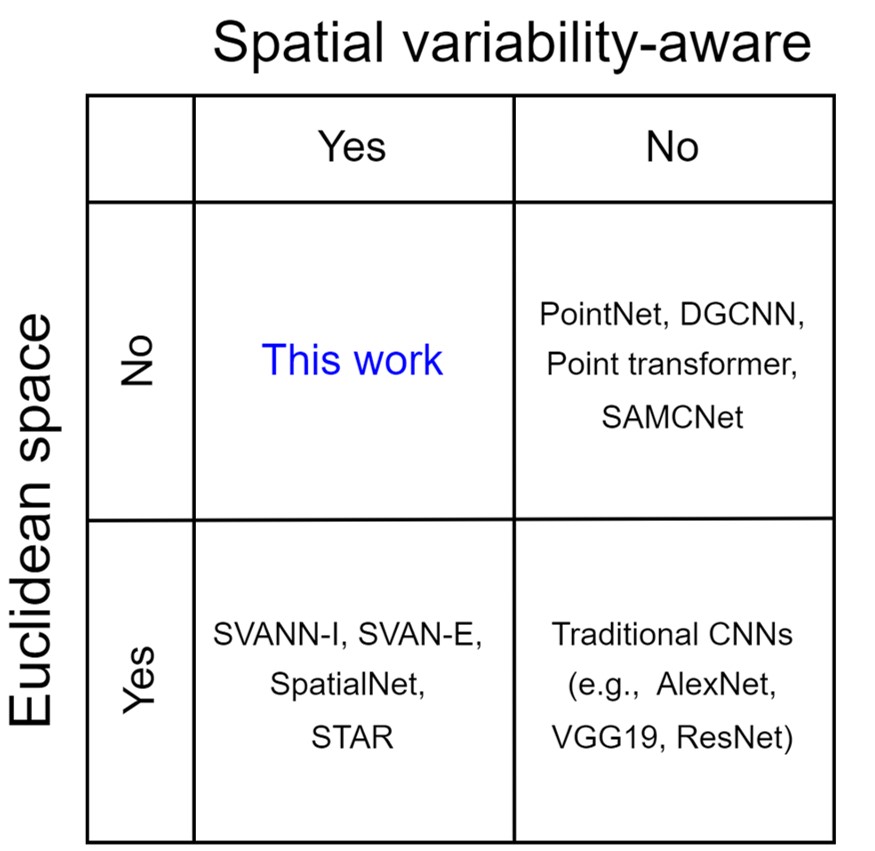}
    \caption{Related works on multi-category point sets for spatially-lucid classification.\\}
    \label{related_work_matrix}
\end{figure}

Research on spatially-lucid classification of multi-category point sets primarily focuses on (1) modeling spatial variability in Euclidean space and (2) representation of point sets using DNNs in non-Euclidean space. Fig. \ref{related_work_matrix} contrasts this with prior related work.\\
\textbf{Spatial Variability in Euclidean Space:} Geographically Weighted Regression (GWR) is a traditional non-parametric model that learns location-specific weighted maps \cite{brunsdon1999some}, but is better suited for linear regression than for complex deep learning prediction tasks. Recent work such as SVANN \cite{gupta2020towards, gupta2021spatial} and \cite{xie2021statistically, xie2021spatial} aims to address spatial variability through deep learning considering location-dependent weights and learning space partitionings of heterogeneous data using weight-sharing mechanisms. However, these methods either require dense training data for each spatial domain and face issues when there is notable spatial variability within a place-type and among similar samples (see Figures \ref{h&e} and \ref{fovs}). For cancer applications, these methods might struggle to capture key spatial interactions, since they assume homogeneous regions \cite{farhadloo2023spatial} and require manual inspection. Notably, these works focus on Euclidean space and overlook spatial variability in \textit{non-Euclidean} spaces like point clouds and cellular maps.\\
\textbf{Point set representation using DNNs:} 
Convolutional neural networks (CNNs) success in pattern recognition \cite{lecun2015deep, he2016deep, cecotti2020grape} has motivated adaptations for direct use on 2D/3D point cloud data, bypassing expensive conversion layers. PointNet \cite{qi2017pointnet} learns the characteristics of the points and aggregates them, while PointNet$++$ \cite{qi2017pointnet++} considers local structures using recursive graph coarsening and PointNet layers. However, these approaches struggle to capture fine-grained local relationships due to permutation invariance. DGCNN \cite{wang2019dynamic} proposes a dynamic graph CNNs by learning PointNet EdgeConv, and constructing local graphs from node-edge features. Techniques based on self-attention networks \cite{zhao2021point,vaswani2017attention} lacks inherent inter-category relationship modeling as it operates on individual point features and focus on whole cloud tasks rather than part relationships. SAMCNet \cite{farhadloo2022samcnet} tackles this by highlighting spatial interactions across diverse point types but don't capture \textit{spatial variability}, favoring scalar over map weights.
\section{Problem Formulation} \label{sec:Problem}
This section reviews several key concepts in spatially-lucid classification and presents the problem statement.

\subsection{Basic concepts}

\begin{figure}[h]
    \centering
    \includegraphics[width=0.85\linewidth]{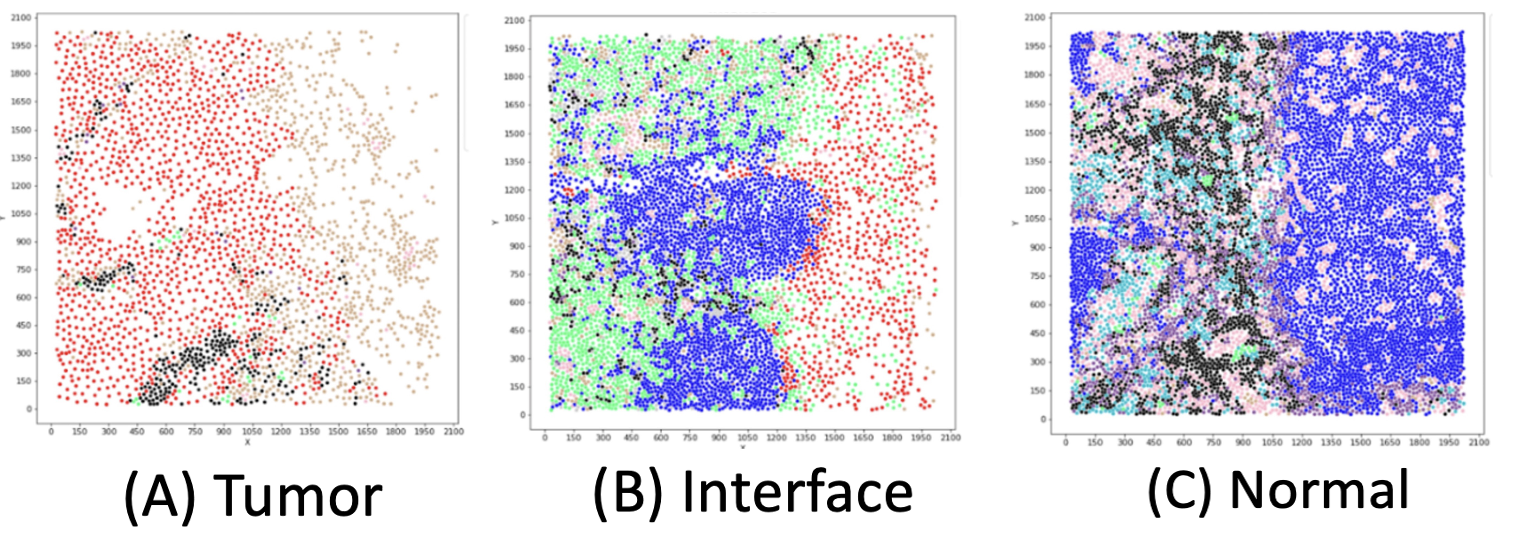}
    \caption{Three multi-category point sets (i.e., pathology-driven fields of view) are categorized based on tumor infiltration. }
    \label{fovs}
\end{figure}

\begin{definition}
A \textbf{spatial domain} refers to a grouping of spatial objects within a specific region in space that share certain characteristics, arrangements, and distribution patterns, distinguishing it from other spatial domains within the same space.
\end{definition}
\begin{definition}
A \textbf{place-type} is a spatial domain $\mathscr{X}$ associated with a probability distribution $P(X)$ over instances $X = \{s_i = (x_i, l_i) | s_i \in \mathscr{X}, i = 1, ..., n\}$, where $x_i$ is a non-spatial categorical attribute, $l_i$ is a two-dimensional vector of spatial point features. The set $X$ forms a \textbf{multi-category point pattern} representing objects (e.g., different cell types) and locations.
\end{definition}
For instance, Fig. \ref{fovs} displays three multi-category point sets from different place types (i.e., tumor, interface, and normal) classified based on tumor infiltration and cell population.

\begin{definition}
A \textbf{spatial arrangement} is the relative positions, or orientations of spatial objects in a space, as well as the patterns and relationships that emerge from their arrangement. 
\end{definition}
For example, co-location patterns \cite{shekhar2001discovering} are an spatial arrangement where subsets of objects frequently occur in close proximity.

\begin{figure}[h]
    \centering
    \includegraphics[width=0.7\linewidth]{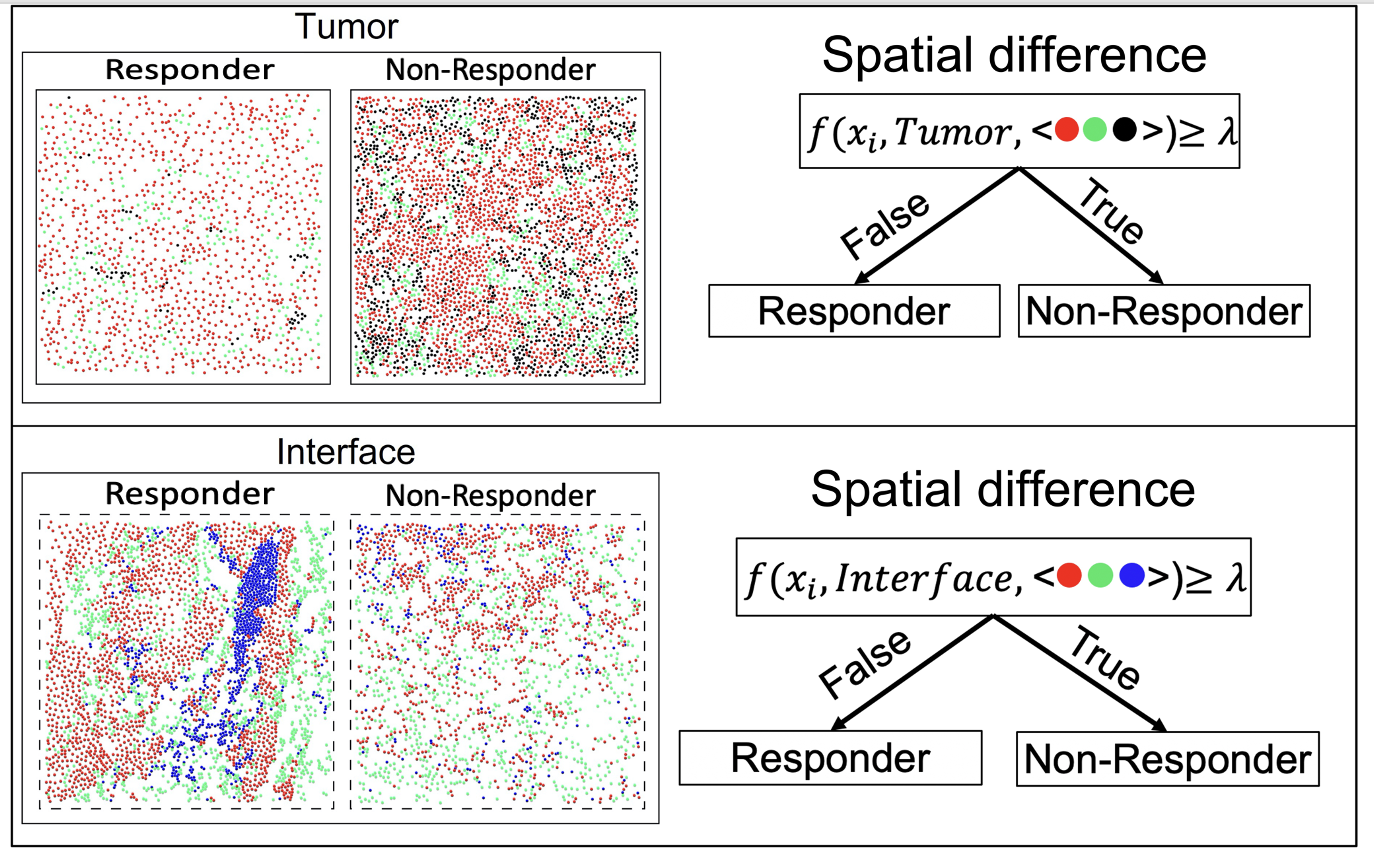}
    \caption{Spatially-lucid classifiers for distinguishing between two class labels belonging to distinct place-types (tumor and interface) based on two types of 3-way spatial arrangements.}
    \label{spatially-lucid}
\end{figure}

\begin{definition}
A \textbf{spatially-explainable classification} $C$ consists of a label space $Y$ and a decision function $f$ that incorporates spatial arrangements, i.e. $C = \{Y, f_A\}$. The decision function $f_A : X \times A \rightarrow Y$ maps from both the instance features $X$ and spatial arrangements $A$ to the label space $Y$. The $f_A$ can be represented as a conditional probability distribution $f_A(x_j, a_j) = \{P(y_k|x_j, a_j)| y \in Y, k = 1, ... , |Y|\}$, where $a_j$ denotes the spatial arrangement associated with instance $x_j$. 
\end{definition}
An example of spatially-explainable classification is shown in Fig. \ref{spatially-lucid} (top row) which separates two class labels (e.g., responder and non-responder) of multi-category point sets in a given place-type (e.g., tumor) based on a 3-way spatial arrangement (<black, green, and red> circles).  

\begin{definition}
\textbf{Spatial variability} refers to the inherent heterogeneity and variation observed in a set of spatial patterns, structures, properties or arrangements in a given spatial domain.
\end{definition}
For example, in Fig. \ref{fovs}, the three selected multi-category point sets (from Fig. \ref{h&e}) depicting place-types (e.g., tumor, interface, and normal regions) illustrate significant spatial variability, revealing variations in cell population across a single tissue sample. 

\begin{definition}
A \textbf{spatially-lucid classification} $\hat{C}$ extends \textbf{spatially-explainable classification} $C$ by incorporating modeling of \textbf{spatial variability} to address inherent heterogeneity in spatial patterns across place-types.

The decision function $\hat{f_A}$ is defined as $\hat{f_A}(x_j, a_j, p_j) = {P(y_k|x_j, a_j, p_j) | y \in Y, k = 1, ... , |Y|}$, where $p_j$ represents the place-type associated with instance $x_j$ and spatial arrangement $a_j$.

The function $\hat{f_A}$ maps instances $x_j$, arrangements $a_j$, and spatial domains $p_j$ to label probabilities, learning both domain-specific arrangement patterns and shared arrangements across domains.
\end{definition}

Figure \ref{spatially-lucid} illustrates the use of a spatially-lucid classifier. In this case, the approach works by learning separate decision functions for each distinct place-type. This technique is effective because it takes into account the inherent spatial variability present across the entire spatial domain, which helps distinguish between the two class labels (i.e., responder and non-responder). Furthermore, these models are explainable because they are based on the spatial arrangements between different data points. This means they can help us identify the most important features that contribute to the classification task. For instance, in the case of tumor classification, important features include relationship between <red, green, and black>.
\begin{definition}
\textbf{Knowledge-guided spatial contextualization} is the use of abstract place-types to represent how objects relate to each other based on their positions.
\end{definition}
For example, in Fig. \ref{h&e}, the normal (blue square) and tumor (red square) place-types exhibit the greatest relative distance, reflecting the tumor's origination from the tissue center and subsequent progression toward the periphery, ultimately infiltrating normal tissue regions and affecting them. The relative distances between place-types are established using expert knowledge (e.g., oncologists). This knowledge guides the learning algorithm during training (Section \ref{sec:ProposedApproach}).

\subsection{Problem Statement}: The problem of spatially-lucid classification can be expressed as follows:\\
\textbf{Input:}
\setlist{nolistsep}
\begin{itemize}[noitemsep]
    \item [--] multi-category point sets from different place-types with class labels 
    \item [--] A distance matrix $D$, where $d_{T_iT_j}$ represents the distance between place-types of type $T_i$ and $T_j$    
    \item [--] A distance threshold, $\alpha$
\end{itemize}
\textbf{Output:} 
\begin{itemize}[noitemsep]
    \item [--] A classifier algorithm for class separation
    \item [--] The most discriminative explanatory features
\end{itemize}
\textbf{Objective:}  Solution Quality (e.g., Accuracy, F1-score)
\textbf{Constraints:} Spatial Variability

\begin{figure}[h]
    \centering
    \includegraphics[width=\linewidth]{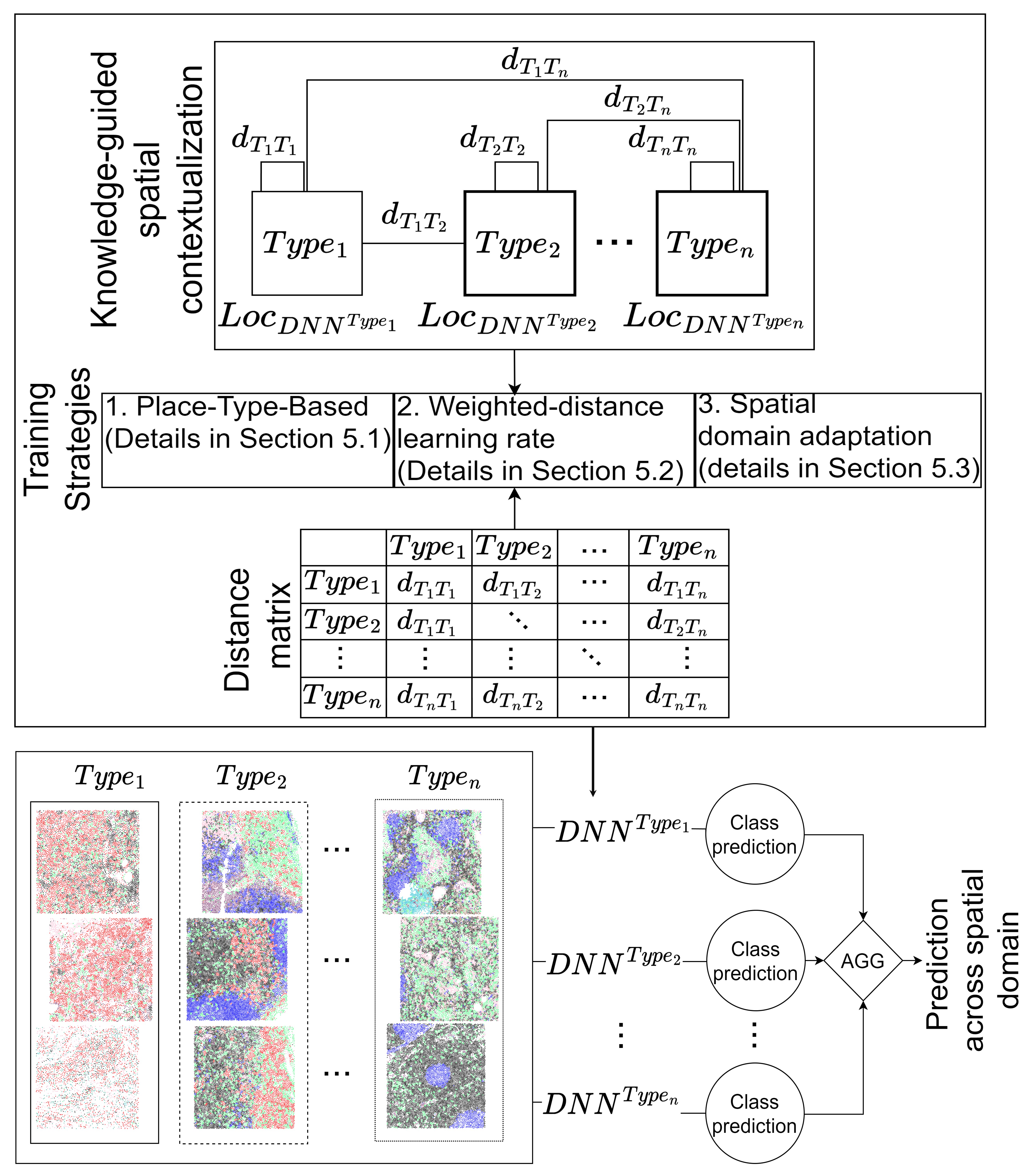}
    \caption{The overall framework of the proposed work.}
    \label{Proposed}
\end{figure}


In this problem, we are given multi-category point sets from various place-types, where each point set is associated with one of two class labels. Additionally, we have a weighted distance matrix, denoted as $D$, where each cell represents the relative distance between two place-types $T_i$ and $T_j$, as defined by domain experts. The objective is to develop a spatially-lucid classifier algorithm that accurately separates the two class labels. Since each place-type belongs to a unique spatial domain, the location of the target deep neural network classifier is significant. This positioning is essential for selecting suitable training samples, which is determined by the distance threshold $\alpha$. It allows the model to learn spatial patterns specific to each place-type, ensuring it accommodates spatial variations and yields dependable classification results.

The overall proposed approach is shown in Fig. \ref{Proposed}. This framework deploys a spatial ensemble technique involving multiple spatial models via a tailored point-wise convolution to capture relationships among different place-types, where each neural network weight is a map weight that varies across spatial domains. This entails aggregating individual model predictions through a function (e.g., weighted average, majority vote), considering the variability in spatial patterns within each distinct place-type. For example, with a distance threshold $\alpha=1$, each place-type has a separate deep neural network (DNN) trained on respective multi-category point sets. Predictions are then aggregated to determine the class label across the entire spatial domain. However, sufficient learning samples may not be available to train separate DNN classifiers. To address this, we explore alternative training strategies like weighted-distance learning rates and spatial domain adaptation, where all samples can train the target classifier (Sections \ref{wdlr}, \ref{sda}).

\section{Proposed Approach}\label{sec:ProposedApproach}
In this section, we explain the proposed spatial ensemble framework for spatially-lucid classification in non-Euclidean space. We hypothesize that a spatial ensemble combining a set of base classifiers will help create stronger predictions and an explainable model. Taking into account spatial variability, we explore different training strategies that range from restrictive to flexible to address the lack of sufficient learning samples.  

\subsection{Place-type-based}: In our baseline method, we focus on the transformation of multi-category point set data into a non-Euclidean graph structure for analysis by a distinct DNN architecture $h$ (e.g., SAMCNet \cite{farhadloo2022samcnet}, DGCNN \cite{wang2019dynamic}). Each unique place-type within our dataset is associated with a specific DNN architecture, emphasizing the role of spatial contextualization. This context is captured through a user-defined distance threshold $\alpha$, which establishes the spatial relationships and proximity boundaries between place-type instances in our model.

The original dataset is represented as a multi-category point set $X = {s_i = (x_i, l_i) | s_i, i = 1, ..., n}$, where $x_i$ denotes categorical features, and $l_i$ represents spatial features, specifically the relative  locations from the left corner of each instance in $X$. To convert this data, we compute an undirected graph $G=(V, E)$, where $V$ represents the vertices (or nodes) and $E$ the edges. This graph is constructed using a $k$-nearest neighbor (KNN) approach, based on the spatial features $l_i$ of each point and a fixed neighborhood distance $d$. The edges $E$ in the graph are determined by the vertices $V$ and their $k$ nearest neighbors and where $E=V\times K$. This process effectively reformats the original Euclidean point set into a non-Euclidean graph structure, a transformation that has been explored in studies such as \cite{bronstein2017geometric}. The resulting graph structure encodes relational information, capturing connections and proximities that extend beyond the traditional Euclidean framework of distances and angles.

The forward-pass embedding $h$, is a function of a unique place-type $p$ and is influenced by a subset of its points, notably the spatial arrangements $a$, is formulated as:
\begin{equation}
\begin{gathered}
    h^{(k+1)}_s(a,p) = \\\sigma \left(W^p_k\sum_{u\in a}\alpha_{s_xu_x} h^{(k)}_{u}(a,p) + B^p_k h^{(k)}_{s}(a,p) \right),
\end{gathered}
\end{equation}
where $h^{(k)}_{s}(a,p)$ is the hidden representation of node $s$ at layer $k$, associated with the spatial arrangement $a$ and a place-type $p$, $\sigma$ is a non-linear activation function, such as LeakyRelu, and both $W^p_k$ and $B^p_k$ are place-type specific trainable matrices. While $W^p_k$ aids in neighborhood aggregation, $B^p_k$ focuses on the hidden representation of the target node itself. Finally, $\alpha_{s_xu_x}$ is the learned categorical pairwise association for nodes within arrangements $a$. 

We can feed these embeddings into any loss function (e.g., cross-entropy) and train the weight parameters using stochastic gradient descent, expressed as follows:

\begin{enumerate}
    \item \textbf{Compute Gradients}:
\begin{equation}
\frac{\partial L}{\partial W^p_k} = \frac{\partial L}{\partial h^{(k+1)}_s(a,p)} \sigma' \left( \cdot \right) \sum_{u \in a}\alpha_{s_xu_x} h^{(k)}_{u}(a,p),
\end{equation}

\begin{equation}
\frac{\partial L}{\partial B^p_k} = \frac{\partial L}{\partial h^{(k+1)}_s(a,p)} \sigma' \left( \cdot \right) h^{(k)}_{s}(a,p), 
\end{equation}

\begin{equation}
\frac{\partial L}{\partial \alpha_{s_xu_x}} = \frac{\partial L}{\partial h^{(k+1)}_s(a,p)} \sigma' \left( \cdot \right) W^p_k h^{(k)}_{u}(a,p).
\end{equation}
\item \textbf{Update Parameters}:

\begin{equation}
W^p_k = W^p_k - \eta \cdot \frac{\partial L}{\partial W^p_k},
\end{equation}

\begin{equation}
B^p_k = B^p_k - \eta \cdot \frac{\partial L}{\partial B^p_k},
\end{equation}

\begin{equation}
\alpha_{su} = \alpha_{su} - \eta \cdot \frac{\partial L}{\partial \alpha_{su}}.
\end{equation}
\end{enumerate}
The term $\sigma'$ denotes the derivative of the activation function with respect to its input. In the case of LeakyReLU, this derivative would be piece-wise linear (i.e., 1 for positive inputs and a small constant for negative inputs).

\subsection{Weighted-distance learning rate: \label{wdlr}} In this approach, all training samples from the various spatial domains, denoted as place-types, are used via a distance-weighted learning rate. Training samples closer to the target model are accorded higher importance than those farther away to account for spatial variability. This prioritization is achieved by adjusting the learning rate using the inverse weighted distance between the target model $h$ and an instance $X$. Both $h$ and $X$ are associated with a specific place-type $p$ according to our knowledge-guided spatial contextualization.

The design decision to adopt a knowledge-guided spatial contextualization arises from the complexities associated with the irregular spatial distribution of place-type instances across multiple learning samples. To illustrate, there might be instances where a place-type mainly clusters in the north-eastern part of one sample but is more evenly dispersed in another. Hence, relying solely on a Euclidean distance metric to gauge the distance between the learning samples and the target model's location could be inadequate. The proposed method not only discerns these inconsistencies and delivers a distance value consistent across diverse learning samples. The method is designed with domain experts (i.e., oncologists) to establish semantics for each place-type and corresponding distances.


We can denote the distance from a target model, associated with a specific place-type $p_i$ and represented by  $h_{p_i}$, to an instance  $X_{p_j}$ from another place-type  $p_j$, as  $d_{h_{p_i}, X_{p_j}}$. The modified learning rate is:
\begin{equation}
\eta_{h_{p_i}, X_{p_j}} = \eta \times \frac{1}d_{h_{p_i}, X_{p_j}},    
\end{equation}
To update the model weights, we need to substitute the original learning rate with $\eta_{h_{p_i}, X_{p_j}}$. Assume we have relative distances between place-types represented as $1 \leq d_{T_iT_j} \leq n$, with a distance threshold set at  $3$, and an initial learning rate of  $10^{-3}$. If a place-type, say  $Type_1$, is associated with a DNN model, learning samples from place-types up to three steps away can be chosen (i.e., $Type_2$ and $Type_3$). Beginning from the place-type $Type_1$, the learning rates, are $10^{-3}$, $5*10^{-4}$, and $3*10^{-4}$, respectively.  

\subsection{Spatial domain adaptation (SDA): \label{sda}} 
Traditional machine learning models rely on the assumption of independent and identically distributed data. However, this assumption is often violated in many spatial real-world scenarios such as cancer research and remote sensing. Domain adaptation (DA) \cite{zhuang2020comprehensive} offers a solution to these distributional discrepancies, operating under the premise that there exists an underlying shared structure or pattern between the source and target domains that can be exploited. The primary objective is to mitigate the distribution shift between source and target domains, leveraging the abundant labeled data from $D_S$ to enhance performance on $D_T$, where sufficient labeled data is lacking.
\begin{figure}[h]
    \centering
    \includegraphics[width=\linewidth]{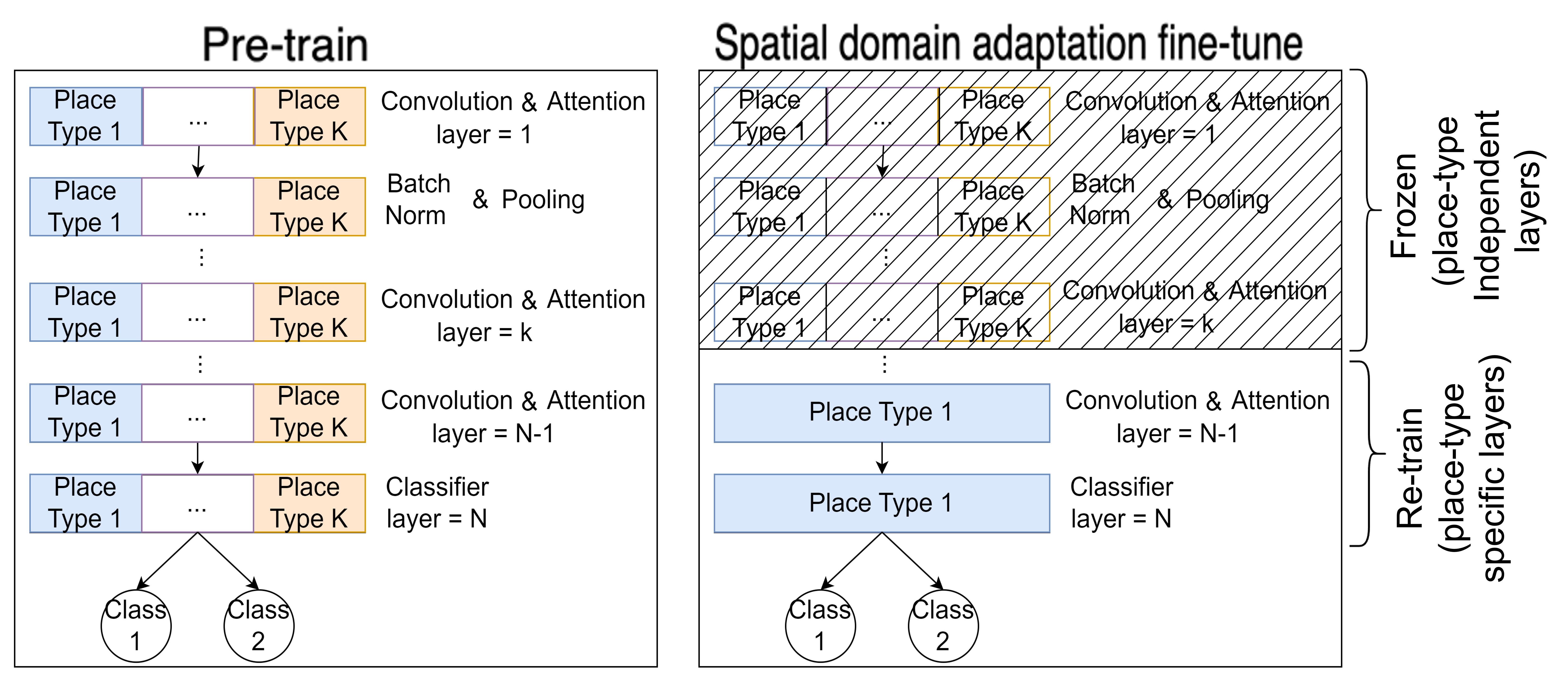}
    \caption{The overall framework of SDA.}
    \label{fig:SDA}
\end{figure}
The spatial domain adaptation strategy, illustrated in Fig. \ref{fig:SDA}, draws inspiration from representation-based domain adaptation (DA) methods. It incorporates a weight-sharing mechanism that categorizes layers into two types: \textit{place-type independent} and \textit{place-type dependent}. The training process consists of two key phases: Initially, a user-specified Deep Neural Network (DNN) architecture (e.g., SAMCNet \cite{farhadloo2022samcnet}, DGCNN \cite{wang2019dynamic}) is pre-trained using multi-category point sets from all place-types. This architecture typically includes $N$ layers, encompassing operations such as convolution, attention, batch normalization, and classification (visible in the left column of Fig. \ref{fig:SDA}). Subsequently, the DNN architecture is split into two segments: the initial $k$ layers, which possess place-type independent parameters, are frozen and shared across all place-types, while the subsequent layers undergo fine-tuning to capture place-type-specific features for the target place-type (e.g., place-type 1 in Fig. \ref{fig:SDA}). The objective function for spatial domain adaptation to the target place-type is expressed as follows:

\begin{equation}
    \begin{split}
    \min_{\theta'} \sum_{j=1}^{n_t} &L(f'(x_{t_j}; \theta', \theta_{k}), y_{t_j}) \\
    &+ \lambda d(f(x_{s_i}; \theta_k), f'(x_{t_j}; \theta'_{k+1:N})), 
    \end{split}
\end{equation}
where, $x_{s_i}$ and $x_{t_j}$ represent instances from the source and target place-types, respectively. $f'$ is the adapted DNN model with parameters $\theta'$ (fine-tuned layers) and $\theta_k$ (frozen layers). $d$ signifies a divergence metric between the source and target domain representations in the shared layers, and $\lambda$ is a hyperparameter that balances the classification loss and domain adaptation loss. $f$ represents the DNN model trained across all place-types. In our experimental setting, we consider a constant $\lambda$ of 1. 
\section{Validation} \label{sec:Experiment} 
\subsection{Experimental Design: \label{setup}} First two experiments were designed to answer \textbf{comparative analysis} while last one for \textbf{sensitivity analysis}:
\begin{itemize}[noitemsep,topsep=0pt]
    \item [1.] Does the proposed method yield better classification performance than competing one-size-fits-all (OSFA) methods? 
    \item [2.]  How does the choice of deep learning architecture for learning spatial relationships affect classification performance?
    \item [3.] What is the impact of number of frozen layers on solution quality?
\end{itemize}

\textbf{Datasets:} The experiments were conducted using a real-world cancer dataset from MxIF images. This dataset consists of three distinct place-types: (1) \textbf{normal} denoted as \textbf{$PT_1$}, (2) \textbf{interface}, denoted as \textbf{$PT_2$}, and (3) \textbf{tumor}, denoted as \textbf{$PT_3$}. The results from each place-type were aggregated to predict class labels across the entire study area. For place-type \textbf{$PT_1$}, the dataset contained 81 multi-category point sets representing two different clinical outcomes of immune therapy. Of these, 38 sets were labeled as responders, while 43 were classified as non-responders, signifying individuals who progressed and experienced tumor recurrence within a year. For place-type \textbf{$PT_2$}, the dataset contained 145 multi-category point sets. Out of these, 68 were identified as responders, with the remaining 77 labeled as non-responders. Lastly, in place-type \textbf{$PT_3$}, out of the 103 point sets provided, 30 were labeled as responders, and 73 as non-responders.

\textbf{Data Preparation:} In each classification task, we divided the data into 80\% training and 20\% testing. Twenty five percent of the training set was selected to be the validation set. Due to the limited number of learning samples, we used data augmentation techniques, including partitioning and rotating the original point set. First, we partitioned the minimum bounding rectangle (MBR) of the point set horizontally by 20\% and 80\% and then 80\% and 20\%. This process ensures that spatial relationship information is kept in each subset and points are not randomly sampled. Next, each learning sample was rotated 16 degrees clockwise three times during the training. We uniformly sampled 1,024 points from each subset for the underlying classification task.

\textbf{Deep Learning  Architectures:} We compared our proposed framework on selected classification metrics with the following state-of-the-art DNN architectures: \textbf{(1) PointNet} \cite{qi2017pointnet++},  a neural network architecture that directly consumes point sets for applications ranging from object classification to part segmentation; \textbf{(2) DGCNN} \cite{wang2019dynamic}, a dynamic graph convolutional neural network architecture for CNN-based high-level point cloud tasks such as classification and segmentation; \textbf{(3) Point Transformer} \cite{zhao2021point}, a DNN architecture that leverages a self-attention mechanism to capture local and global dependencies as well as point cloud tasks such as classification and segmentation; and \textbf{(4) SAMCNet}, a spatial-interaction aware multi-category deep neural network \cite{farhadloo2022samcnet}, for learning N-way spatial relationships in multi-category point sets. All hyper-parameters were tuned through tuning on the validation set. 

\textbf{Evaluation Metric \& Platform:} Model performance was measured via weighted average of accuracy, precision, recall, and F1-score. We used K40 GPU composed of 40 Haswell Xeon E5-2680 v3 nodes. Each node had 128 GB of RAM and NVidia Tesla K40m GPUs, each of which had 11 GB of RAM and 2880 CUDA cores.

\textbf{Candidate methods:} Baseline [B] and Proposed [P] methods are as follows:
\begin{itemize}
    \item \textbf{[B1]} A \textbf{one-size-fits-all (OSFA)} approach, where a single DNN is trained on the entire dataset with no consideration for spatial variability.
    \item \textbf{[P1]}  A proposed \textbf{place-type-based} approach, where a separate DNN model is trained for each place-type (i.e., normal, interface, and tumor). 
    \item \textbf{[P2]} A proposed \textbf{weighted-distance learning rate} approach, where a DNN model is trained across all place-types, while samples closer to target model have higher learning rate than ones farther away.
    \item \textbf{[P3]} A proposed \textbf{spatial-domain-adaptation}, where a DNN model is pre-trained across all place-types, and then it is fine-tune for the target-place-type. 
\end{itemize} 

\subsection{Experiment Results}: This section presents the results of our spatially-lucid classification assessment.

\emph{\textbf{Comparative Analysis:}}
We conducted experiments to evaluate candidate DNN architectures on the classification tasks described in Section \ref{setup}. The results are summarized in Table \ref{tab:comp}. Our findings indicate that the proposed spatial ensemble framework, incorporating place-type-based [P1] and spatial-domain-adaptation [P3] approaches, demonstrates better classification performance compared to the OSFA [B1] setting in the majority of cases. Notably, PointNet (Table \ref{tab:pointnet}) with a weighted-distance learning rate [P2] improves classification accuracy by 7\%. Point Transformer with spatial-domain-adaptation consistently outperformed OSFA in classification accuracy by a margin of 7\%. Lastly, SAMCNet with single-place-type [P1] and spatial domain adaptation [P3] approaches consistently outperformed OSFA setting accuracy by 14\% and 11\%, respectively.

\begin{table}[H]\scriptsize
\caption{Classification performance across candidates.}
\label{tab:comp}

\begin{subtable}[t]{0.4\textwidth}
\centering
\begin{tabular}{|l|l|l|l|l|}
\hline
Baseline & Accuracy & F1-Score & Precision & Recall \\ \hline
[B1.] & 0.571 & 0.416 & 0.327 & 0.571 \\ \hline
[P1.] & 0.429 & 0.405 & 0.457 & 0.429 \\ \hline
\textbf{[P2.]} & \textbf{0.643} & \textbf{0.626} & \textbf{0.675} & \textbf{0.643} \\ \hline
[P3.] & 0.357 & 0.347 & 0.34 & 0.357 \\ \hline
\end{tabular}
\caption{Competition 1: Pointnet \cite{qi2017pointnet}.}
\label{tab:pointnet}
\end{subtable}
\hfill
\begin{subtable}[t]{0.4\textwidth}
\centering
\begin{tabular}{|l|l|l|l|l|}
\hline
Baseline & Accuracy & F1-Score & Precision & Recall \\ \hline
{[}B1{]} & 0.571 & 0.514 & 0.571 & 0.571 \\ \hline
\textbf{{[}P1{]}} & \textbf{0.572} & \textbf{0.533} & \textbf{0.606} & \textbf{0.572} \\ \hline
{[}P2{]} & 0.542 & 0.459 & 0.575 & 0.50 \\ \hline
{[}P3{]} & 0.50 & 0.381 & 0.25 & 0.50 \\ \hline
\end{tabular}
\caption{Competition 2: DGCNN\cite{wang2019dynamic}}
\label{tab:dgcnn}
    \end{subtable} \\
    \hfill
    \begin{subtable}[t]{0.4\textwidth}
\begin{tabular}{|l|l|l|l|l|}
\hline
Baseline & Accuracy & F1-Score & Precision & Recall \\ \hline
{[}B1{]} & 0.50 & 0.46 & 0.576 & 0.50 \\ \hline
{[}P1{]} & 0.50 & 0.381 & 0.307 & 0.50 \\ \hline
{[}P2{]} & 0.55 & 0.479 & 0.464 & 0.428 \\ \hline
\textbf{{[}P3{]}} & \textbf{0.572} & \textbf{0.533} & \textbf{0.606} & \textbf{0.571} \\ \hline
\end{tabular}
\caption{Competition 3: Point Transformer \cite{zhao2021point}}
\label{tab:pointTransformer}
\end{subtable}

\begin{subtable}[t]{0.4\textwidth}
\centering
\begin{tabular}{|l|l|l|l|l|}
\hline
Baseline & Accuracy & F1-Score & Precision & Recall \\ \hline
{[}B1{]} & 0.714 & 0.714 & 0.714 & 0.714 \\ \hline
\textbf{{[}P1{]}} & \textbf{0.857} & \textbf{0.857} & \textbf{0.857} & \textbf{0.857} \\ \hline
{[}P2{]} & \textbf{0.806} & \textbf{0.806} & \textbf{0.829} & \textbf{0.826} \\ \hline
\textbf{{[}P3{]}} & \textbf{0.824} & \textbf{0.856} & \textbf{0.869} & \textbf{0.824} \\ \hline
\end{tabular}
\caption{Competition 4: SAMCNet \cite{farhadloo2022samcnet}}
\label{tab:samcnet}
\end{subtable}
\end{table}
Most importantly, it can be observed that the choice of DNN architecture may play a significant role due to the importance of learning spatial relationships in multi-category point patterns. SAMCNet (Table \ref{tab:samcnet}) outperforms all other competitors significantly. Most notably, our proposed spatial ensemble framework with SAMCNet was able to improve accuracy, F1-score, precision, and recall over all other state-of-the-art DNN architectures by a margin of 33.6\%, 39.6\%, 37.2\%, and 34.1\%, respectively.

\begin{figure}
    \centering
    \includegraphics[width=0.8\linewidth]{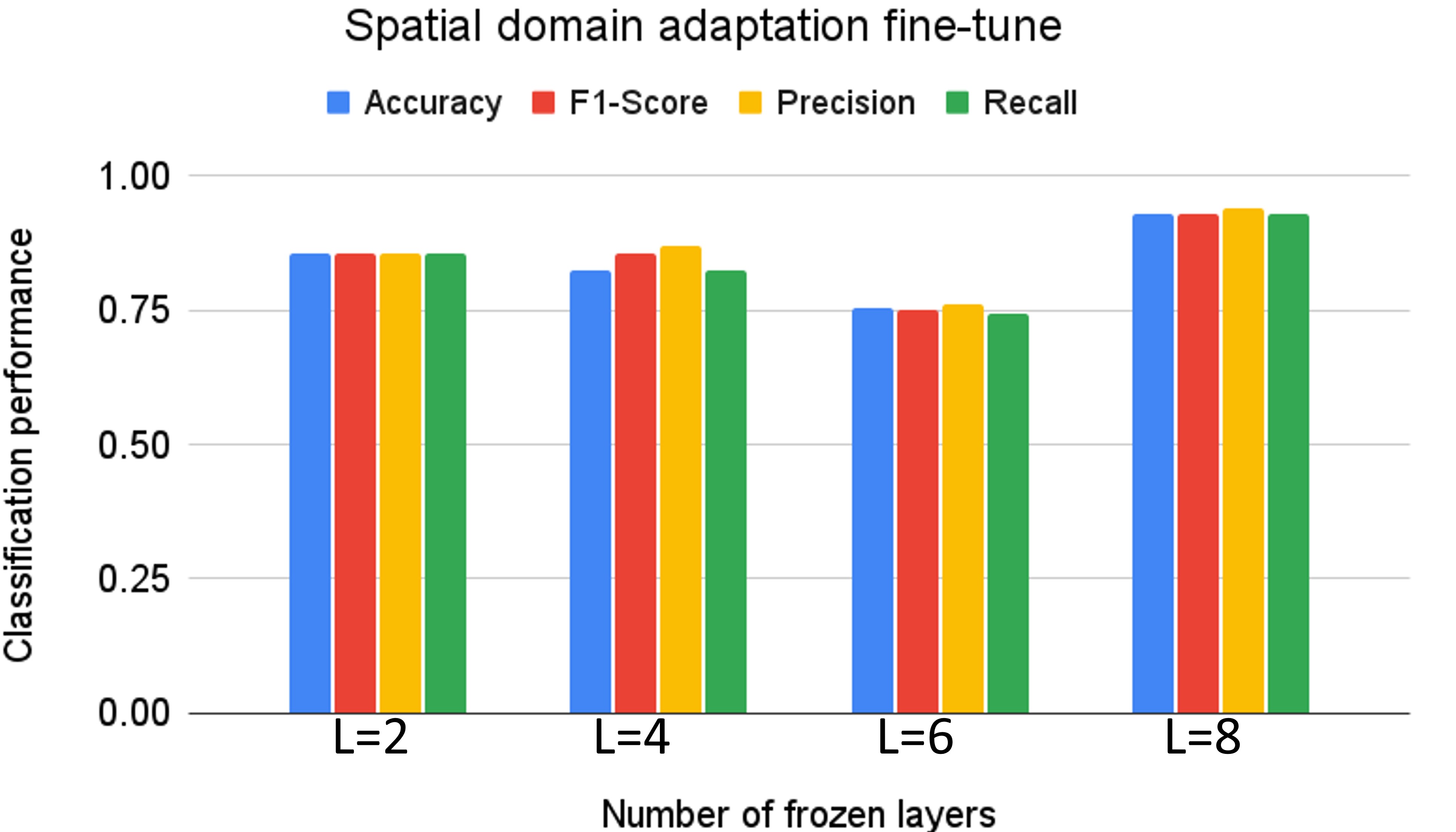}
    \caption{Solution quality based on number of frozen layer in spatial domain adaptation.}
    \label{sens:I}
\end{figure}

\emph{\textbf{Sensitivity Analysis}}\textbf{:} We performed a sensitivity analysis to evaluate the influence of number of frozen layers in spatial domain adaptation [P3] with best DNN candidate method (i.e., SAMCNet).\\ 
\textbf{Impact of the number of frozen layers in [P3]:}
Figure \ref{sens:I} displays the classification performance as the number of frozen layers increases during the fine-tuning of the model for place-type dependent layers. It is evident that there exists a trade-off between re-training the majority of the network parameters for the target place-type and classification performance. This finding highlights that fine-tuning the network on the target place type improves classification performance compared to the OSFA setting. 

Contrary to the initial belief that retraining the majority of parameters would lead to better classification, our findings, as demonstrated by only two frozen layers, suggest otherwise. However, as we increase the number of frozen layers, with an exception at layer 6, the classification performance increases. This is in line with the hypothesis that there is an underlying shared structure across place-types that assists in improving performance for the target-place-type.

\section{Case Study}\label{case-study}
We conducted a case study whose objective was to compare the intracellular interactions obtained by using a single DNN model for an entire tissue sample (OSFA \textbf{ [B1]}) versus using multiple models (\textbf{[P1]}), where each model is dedicated to a specific place-type (e.g., interface or tumor), to classify the input sample as either responder or non-responder. For this purpose, we used a trained SAMCNet \cite{farhadloo2022samcnet} DNN to extract features after the point pair prioritization network at layer-4, where the model has learned spatial and prioritization associations.

We evaluated the importance of the identified spatial relationships, namely, the SAMCNet representation of subsets of point types, through permutation feature importance. This metric measures the importance of a feature by evaluating the decrease in model performance by randomly shuffling the feature space. Previously, we have demonstrated the effectiveness of interpretable models that utilize hand-constructed spatial quantification methods (e.g., participation ratio \cite{shekhar2001discovering}), coupled with decision tree algorithms \cite{li2021srnet}. The top three most relevant spatial associations found within the interface, tumor, and entire tissue sample are shown in Tables \ref{import_osfa}, \ref{import_core}, \ref{import_inter}, respectively. Following is a brief interpretation of these results from a clinical standpoint.

\textbf{Clinical Implications:} 
In Table \ref{import_osfa}, interpreting biological significance is challenging. The spatial relationship between helper T cells and macrophages denotes TH1 subtype's role in macrophage activation, possibly influencing ICI therapy response. Angiogenesis intensity, indicated by vasculature cells' spatial relationship, is linked to tumors ensuring nutrient supply. Overall, the results in Table \ref{import_osfa} capture spatial relationships characteristic of various biological dynamics occurring in metastatic tissue, rather than being specific to a particular place-type. 
 
\begin{table}[tbh!]
\scriptsize
\caption{The most relevant spatial relationships in entire tissue using the OSFA approach.}
\begin{tabular}{|c|c|c|}
\hline
Rank & Center cell & Neighboring cells \\ \hline
1    & Vasculature & Helper T cell, Macrophage, Vasculature  \\ \hline
2    & Helper T cell & \begin{tabular}[c]{@{}c@{}}Helper T cell, Macrophage, Tumor cell, \\ Vasculature\end{tabular} \\ \hline
3    & Macrophage    & Helper T cell, Macrophage \\ \hline
\end{tabular}
\label{import_osfa}
\end{table}

Table \ref{import_core} shows key spatial relationships within the tumor place-type involving tumor cells, macrophages, and vasculature. Macrophages' interactions highlight their role in engulfing damaged cells, crucial for the tumor environment. The blood supply's significance in tumor growth is underscored by the tumor cells' spatial relation with vasculature. Therefore, these intracelluar interactions are very relevant specifically for the tumor place-type.

\begin{table}[h!]
\scriptsize
\caption{The most relevant spatial relationships in the tumor place using the place-type-based \textbf{[P1]} approach.}
\begin{tabular}{|c|c|c|}
\hline
Rank & Center cell & Neighboring cells                                                \\ \hline
1    & Tumor cell & Tumor cell, Vasculature \\ \hline
2    & Macrophage & Macrophage, Tumor cell                \\ \hline
3    & Tumor cell & Macrophage, Tumor cell, Vasculature           \\ \hline
\end{tabular}
\label{import_core}
\end{table}

Table \ref{import_inter} illustrates notable spatial patterns in the interface place-type. B cells have significant interactions with helper T cells, tumor cells, and macrophages, emphasizing their role in antibody production and immune responses against cancer, especially at the tumor-lymph node boundary (i.e, interface place-type). Helper T cells' presence aids B cell activation and assist in cytotoxic T cell functions against cancer cells. The intensity of this process may vary between responders and non-responders.

\begin{table}[tbh!]
\scriptsize

\caption{The most relevant spatial relationships in the interface place using the single-place-type \textbf{[P1]} approach.}
\begin{tabular}{|c|c|c|}
\hline
Rank & Center cell & Neighboring cells                                                          \\ \hline
1    & B cell & B cell, Helper T cell, Vasculature                           \\ \hline
2    & B cell & B cell, Helper T cell, Tumor cell                             \\ \hline
3    & B cell & B cell, Macrophage, Regulatory T cell \\ \hline
\end{tabular}
\label{import_inter}
\end{table}

Understanding cellular interactions enhances the development of potent cancer immunotherapies. The place-type specific observations offer insights on tumor progression and ICI therapy response through place-type dependent patterns, yielding precise therapeutic insights. In contrast, the OSFA settings provide place-type independent results, which, while broader, might be more challenging to directly associate with specific ICI therapy outcomes.
\section{Conclusion \& Future Work} \label{sec:Conclusion}

We investigated a spatially-lucid classification deep neural network for multi-category point sets in non-Euclidean space. Our approach introduces a spatial ensemble framework where network parameters vary as a map across place-types, in contrast to the scalar parameters used in traditional OSFA. Additionally, we introduced flexible training strategies that leverage samples from all place-types, addressing challenges related to insufficient training data. Experiments show that the proposed model outperforms existing DNN techniques.

For future work, we aim to delve deeper into spatial interpretability and variability within individual place-types, focusing on the density and distribution of spatial interactions in different sub-regions. We also intend to adapt generative models, like GANs \cite{goodfellow2020generative, zhuang2020comprehensive}, to spatial data, aiming to capture a broader range of spatial interactions, from place-type independent to place-type dependent.

\section*{Acknowledgments}{This material is based on work supported by the NSF under Grants Nos. 1901099, and 1916518; and the USDA under Grant Nos. 2023-67021-39829 and 2021-51181-35861. We also thank Kim Koffolt and Spatial Computing Research Group for their valuable comments and refinements.}

\bibliographystyle{unsrt}
\bibliography{citation}

\clearpage

\end{document}